# Current Tomography - Localization of void fractions in conducting liquids by measuring the induced magnetic flux density

L. Krause[1*], N. Kumar[2], T. Wondrak[1], S. Gumhold[2], S. Eckert[1], K. Eckert[1,3]

[1]Helmholtz-Zentrum Dresden-Rossendorf, Bautzner Landstraße 400, 01328 Dresden, Germany

[2]Computer Graphics and Visualization, Faculty of Computer Science, Technische Universität Dresden, Germany

[3]Transport Processes at Interfaces, Institute of Process Engineering and Environmental Technology, Technische Universität Dresden, Germany

*Email: l.krause@hzdr.de

## ABSTRACT

*A novel concept of a measurement technology for the localization and determination of the size of gas bubbles is presented, which is intended to contribute to a further understanding of the dynamics of efficiency-reducing gas bubbles in electrolyzers. A simplified proof-of-concept (POC) model is used to numerically simulate the electric current flow through materials with significant differences in electrical conductivity. Through an automated approach, an extensive data set of electric current density and conductivity distributions is generated, complemented with determined magnetic flux densities in the surroundings of the POC cell at virtual sensor positions. The generated data set serves as testing data for various reconstruction approaches. Based on the measurable magnetic flux density, solving Biot-Savart's law inversely is demonstrated and discussed with a model-based solution of an optimization problem, of which the gas bubble locations are derived.*

## 1   INTRODUCTION

A transition away from greenhouse gas-emitting processes is essential to minimize the impacts of climate change. Using electricity from renewable sources, the electrolytically produced energy carrier hydrogen satisfies this demand. By applying an electrical current to water, the hydrogen evolution reaction (HER) and oxygen evolution reaction (OER) are initiated, and $H_2O$ reacts to $H_2$ and $O_2$ at the electrode surfaces. The electrically non-conductive gaseous products of water electrolysis are desired, but their accumulation in the vicinity of the electrodes impede the current flow, block reaction sites on the electrocatalysts and thus reduce the efficiency of the electrolyzer (Angulo et al., 2020).

In compact industrial water electrolyzer stacks, evolving gas bubbles are optically not accessible. For the purpose of investigating bubble nucleation, better understanding bubble dynamics and removing gaseous blockages under these conditions, non-optical measurement techniques are needed to localize $H_2$ and $O_2$ bubbles (Yuan et al., 2023), such as neutron imaging (Panchenko et al., 2018), X-ray computed tomography (Satjaritanun et al., 2020) or frequency analysis of current fluctuations (Kwan et al., 2022).

With the application of the cell voltage to the electrolyzer, an electric current starts to flow. As a result, a magnetic field is induced in the surroundings of the electrolytic cell, which is related to the current by means of the Biot-Savart law. Solving the equation inversely, Hauer et al. have demonstrated the possibility of reconstructing the current density distribution in fuel cells from the non-invasively measured magnetic flux density (Hauer et al., 2005). In combination with a highly conductive electrolyte and gas bubbles of negligibly low electrical conductivity, bubbles are influencing the current distribution, like shown in Figure 1. Alterations of the current density affect the induced magnetic field. Solving the inverse problem of the Biot-Savart law for a water electrolyzer, possibly both the bubble distribution can be estimated from the reconstructed current density, and undesired inhomogeneities of the current density can be detected.

Within the scope of this study, we aim to develop a model to prove the concept of localizing gas bubbles and estimating their sizes by reconstructing the current density and the conductivity based on measurements of the magnetic flux density. The numerically simulated model simplifies the electrolyzer as a reaction-free electrical conductor, consisting of conducting and non-conducting materials. Combined with calculated magnetic flux densities at virtual sensor positions, a data set will be generated in an automated process to serve as testing data for several reconstruction approaches. A model-based solution of an optimization problem demonstrates the inverse solution of Biot-Savart's law, and hence the localization of void fractions with negligibly low conductivity.

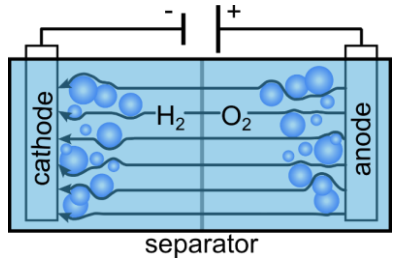

**Figure 1. Current distribution in a water electrolysis cell.**

## 2   METHODS

### 2.1   Proof-of-concept model

Proving the possibility of localizing and measuring non-conducting bubbles by reconstructing the current density and conductivity distribution based on the measured surrounding induced magnetic flux density, a numerically studied proof-of-concept model (POC) is developed. The POC model generates testing data for different conductivity and current density reconstruction methods, such as a Neural Network (NN) based approach presented Kumar et al. (2023) and a model-based classical solution of an optimization problem.

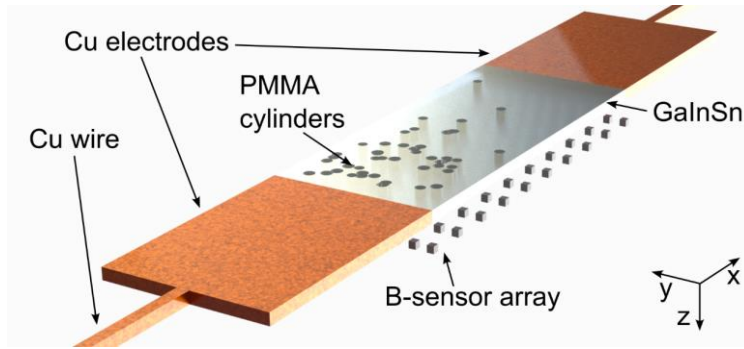

**Figure 2. Simulated setup.**

For this purpose, the water electrolyzer is simplified to a quasi-two-dimensional setup, consisting of materials with large conductivity differences. The dimensions of the numerically studied setup are chosen with consideration of an experimental validation to come. A channel (length x width x height = 16 x 7 x 0.5 cm), shown in Figure 2, is filled with liquid GaInSn. An electric current is applied through two Cu electrodes, each with a length x width x height of 10 x 7 x 0.5 cm. Anode and cathode are contacted by wires, modelled with lengths of 50 cm and quadratic cross-sections with side lengths of 0.5 cm. Enabling the automatic generation of large data sets as NN training data, various geometrical configurations of the model, meaning diverse locations and sizes of regions with less conductivity, are compiled from a java-class file in the finite element software COMSOL Multiphysics V6.0 (COMSOL Inc, Burlington, USA). Therefore, between 30 and 120 PMMA cylinders with radii $R_{cyl}$ = 2 … 2.5 mm are placed in the liquid metal. The cylinder sizes are oriented on bubble agglomerates. The lower constraint of the range was chosen to limit the minimal mesh cell sizes to decrease computation time for the generation of large data sets and with respect to the spatial current density and conductivity resolution for the NN training data, discussed later. Larger bubble clusters are represented by merged cylinders. The locations of the cylinder x-coordinates are normal distributed after $x = Z_x \sigma_x + \mu_x$ with mean values of -7.5 cm ≤ $\mu_x$ ≤ 7.5 cm. Here $Z_x$ is a random variable distributed according to a normal distribution with standard deviation of one. This is stretched by $\sigma_x$ sampled uniformly in the range of [2 cm, 5 cm]. The y-coordinates are uniformly distributed from [-3.5 cm, 3.5 cm]. Since no electrochemical reactions occur in the liquid metal after the application of a current, concentration-induced conductivity gradients are excluded. A low electric conductivity $\sigma$ of $5 \cdot 10^{-14}$ S/m simulates the void fractions at positions of PMMA cylinders (Zhang et al., 2015). For the Cu wires and electrodes $5.8 \cdot 10^7$ S/m and for GaInSn $3.3 \cdot 10^6$ S/m are used (Plevachuk et al., 2014). A current density of 1 A/cm$^2$ is applied at the electrode surface facing the liquid metal with edge lengths of 7 cm x 0.5 cm. The current density is thus between the typical values of alkaline and PEM electrolyzers. Since the copper wire with a smaller cross-section conducts the input current, 14 A/cm$^2$ need to be applied corresponding to a current of 3.5 A.

As the current density simulations are performed in 3D to investigate all spatial components of the magnetic flux density, the magnetic field is calculated by solving Biot-Savart's law only at the positions of virtual sensors to minimize the computation time. In a x-y-plane 10 x 10 and 50 x 50 sensors are positioned. The virtual sensor array is located 3, 5, 10 and 25 mm below the electric current-carrying part containing GaInSn.

## 2.2 Physical background

Since for simplification of the electrolyzer neither the liquid metal nor the poorly conducting void fractions are moving, the electric field at a position *r'* in 3D space

$$\boldsymbol{E}(\boldsymbol{r'}) = -\nabla(\varphi(\boldsymbol{r'})) \qquad (1)$$

is defined as the negative gradient of the electric potential $\varphi(r')$. Multiplied with the electric conductivity $\sigma(r')$, Ohm's law is solved under stationary conditions to calculate the current density

$$\boldsymbol{j}(\boldsymbol{r'}) = -\sigma(\boldsymbol{r'})\nabla(\varphi(\boldsymbol{r'})) \qquad (2)$$

that is considered as divergence-free $\nabla \cdot j(r') = 0$. According to Biot-Savart's law the magnetic flux density

$$\boldsymbol{B}(\boldsymbol{r}) = \frac{\mu_0}{4\pi} \iiint_V \frac{\boldsymbol{j}(\boldsymbol{r'}) \times (\boldsymbol{r}-\boldsymbol{r'})}{|\boldsymbol{r}-\boldsymbol{r'}|^3} dV' \qquad (3)$$

is induced at a position *r* outside the conductor. $\mu_0$ of $4\pi \cdot 10^7$ N/A$^2$ denotes the magnetic permeability of the vacuum and *dV'* the volume element of the current density *j(r')* with the integration variable *r'*.

## 2.3 Simulation parameters and mesh transformation

The current density distribution was simulated numerically using second-order shape functions. To enable an automated grid generation for various geometries, the geometry is divided in finite tetrahedral elements into an unstructured mesh. After a grid independency study, the mesh was refined in regions with high current density gradients, such as the wire – electrode interfaces and the GaInSn containing volume. For the liquid metal, a minimal element size of 0.1 mm and a maximal element size of 5 mm was chosen with additional increased resolution of narrow regions, since this area is to be reconstructed and mainly the non-conducting PMMA cylinders affect the current density distribution.

Further processing of the simulated current density and conductivity distributions and calculation of *B(r)* fields allows their use for several reconstruction approaches and is schematically shown in Figure 3. The *j(r')* and *σ(r')* computation of multiple geometries requires meshes with different numbers of cells. Since NNs need defined dimensions of input arrays, the initially tetrahedral mesh is transformed to a grid of hexahedrons with a fixed element number. The current density distribution of the structured mesh with one cell layer in height can be considered as 2-dimensional, since the z-component and variation of x- and y-components along z-direction of the current are negligible. The grid, consisting of a total of 774 cells, is higher resolved in the GaInSn volume with 510 near-cubic cells, each with length x width x height = 4.71 x 4.67 x 5 mm. The current density and electrical conductivity inside each hexahedron are computed by using inverse distance weighted interpolation (Shepard, 1968) of the 24 nearest tetrahedrons.

Subsequently, the magnetic flux density distribution is calculated using Biot-Savart's law (Eq. 3) at the position of the virtual sensors based on $j_{tet}(r')$ and $j_{tet}(r')$. Since only one spatial component of $B(r)$ will be measurable in a planned experimental setup for validation, further reconstruction is based on the $B_x(r)$ vector that contains the most information.

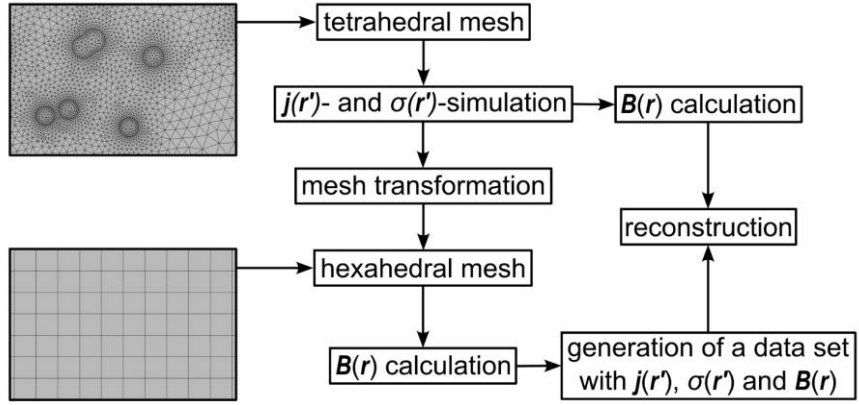

**Figure 3. Post-processing of simulated data.**

Finally, the magnetic flux density is used for the reconstruction. In the scope of this study, a model-based solution of an optimization of current streamline paths is presented that focuses on assessing cylinder positions from $B_{x,tet}(r)$. Streamlines represent the paths of divergence-free currents through a conductor and are always tangential to the current. Path deviations, caused by obstacles like small electric conductivities, lead to a change of the line density, which implies a modification of the current density and can be detected in fluctuations of the magnetic flux density. Representing current streamlines, 15 polygonal chains, each with 3.5/15 A, are created within the dimension of the liquid metal volume in Python. The pathway of each of the polygonal chains is defined by 40 vertices along the x-axis, initially without path deviations along an y-coordinate, parallel and equally spaced to each other. By solving Biot-Savart's law the magnetic flux density $B_{x,poly}(r)$ is calculated, based on the streamline paths. $B_{x,poly}(r)$ is subtracted from the simulated reference field. The sum of the squared absolute differences

$$B_{x,\text{diff}} = \sum_{i=1}^{n_\text{sensor}} \left| B_{x,\text{tet}}(r) - B_{x,\text{poly}}(r) \right|^2 \tag{4}$$

serves as function to be reduced. Using an optimization function (sequential least squares programming, SLSQP) $B_{x,\text{diff}}$ is minimized by alternating the y-coordinates of the streamlines iteratively. The convergence criterion was set to $10^{-10}$. Further approaches to solve the inverse problem and additionally estimating void fraction sizes based on $B_{x,\text{hex}}(r)$ are discussed in Kumar et al. (2023).

## 3  RESULTS AND DISCUSSION

As a result of the study, a data set of current density and conductivity distributions of 10,000 geometrical configurations was simulated in a tetrahedral mesh, exemplified as current density distribution and streamlines of the GaInSn containing region in Figure 4A (left). After the transformation to the hexahedral mesh (Figure 4A, right), the positions of the cylindrical void fractions are still visible in the coarser grid as dark blue fields surrounded by maxima of the current density. For the reconstruction methods in Kumar et al. (2023), $B_x(r)$ of a 10 x 10 sensor array with 5 and 25 mm distance to the conductor was calculated for the whole set using the hexahedral meshed $j(r')$.

In this study, the configuration in Figure 4A was exemplarily used for the reconstruction of void fraction localizations by optimizing streamline paths. Figure 4B shows the $B_x(r)$ fields calculated from the tetrahedral grid for 10 x 10 and 50 x 50 virtual sensors with distances of 3, 5 and 10 mm. The displayed magnetic flux densities are located at the centers of the cells, demonstrating the array resolutions. Comparing 50 x 50 and 10 x 10 sensors at 3 mm distance, the lower sensor resolution causes a spatially smoothed magnetic field. Simultaneously, the amplitude and the degree of details of $B_x(r)$ decrease with an increasing sensor-conductor gap. At a distance of 10 mm, the magnetic field has already decayed to an extent that the local extrema, as well as other characteristics, are similar for both array configurations.

Further, the path optimization approach simplifies the current density distribution to streamlines. Polygonal chains closer to each other indicate an increased current density *j(r')*. The reconstructed streamlines in Figure 4C belong to the $B_x(r)$ distributions in Figure 4B. First, the results of the virtual sensor arrays in 3 and 5 mm distances are compared with the simulated *j(r')* reference (Figure 4A, left) with focus on the three colored boxes. Larger obstacles, like the connected cluster of cylinders underneath the blue window, are visible for both sensor configurations as downward bent polygonal chains above the void fraction. The squeezed streamlines in the red box are barely noticeable with 10 x 10 sensors 5 mm away from the cell. Due to the reduced spatial resolution of the virtual sensor configuration in combination with the increased distance, local extrema of $B_x(r)$ are insufficiently detected. Likewise, small magnetic flux density extrema are not resolved properly with 10 x 10 sensors in the region marked in green surrounding the single cylinder. The $B_x(r)$ field smoothing inhibits the streamline path reconstruction in these regions. For the sensor distance of 10 mm, essential details of the $B_x(r)$ field are weakly pronounced, and the paths of the streamlines is insufficiently reconstructed. Whether the reconstruction is possible depends on the number of streamlines and the convergence criterion of the optimization function. In this study, a small number of polygonal chains was chosen for the sake of clarity. Nevertheless, the number of sensors must be reasonably high and their distance to the current-conducting plane has to be as small as possible. Therefore, tiny sensors are needed, such as those developed by Schmidtpeter et al. (2023).

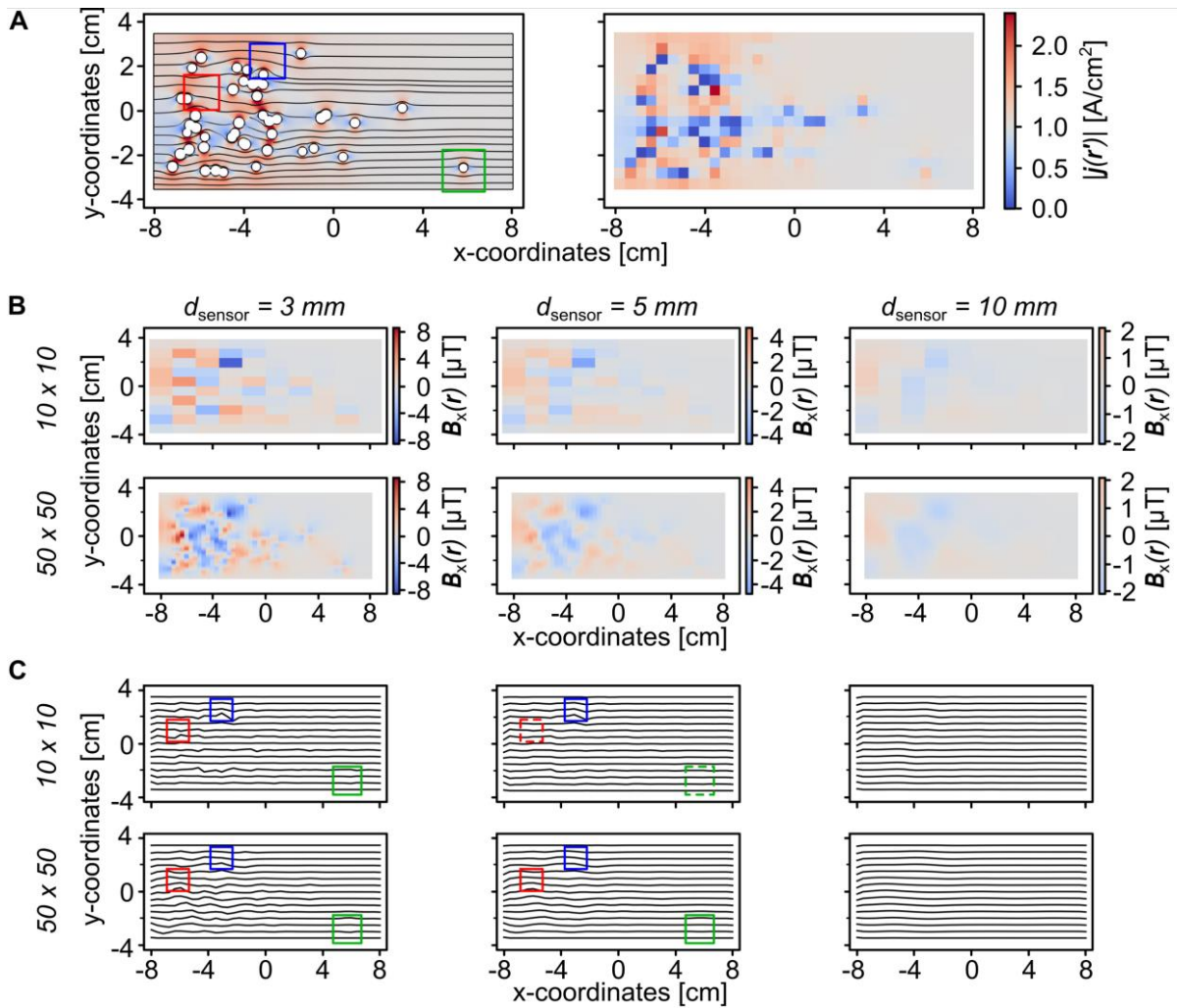

**Figure 4. Results of the simulations. (A) Current density distribution in the tetrahedral (left) and the hexahedral mesh (right). (B) x-component of the magnetic flux density for 10 x 10 and 50 x 50 virtual sensors with distances of 3, 5 and 10 mm between conductive and sensor plane. (C) Reconstructed streamlines based on the $B_x$-results in (B).**

## 4 CONCLUSION AND OUTLOOK

The concept of a novel non-invasive measurement technique for the localization and size estimation of non-conductive fractions in current-conducting liquids was presented in this study. Predictions are made by reconstructing the current density and conductivity distribution based on the induced magnetic flux density. The developed model simplifies a water electrolyzer to a conductor with strong gradients of electrical conductivity. In an automated process, the current density and conductivity distributions were numerically simulated for different geometric variations of PMMA arrangements in liquid GaInSn. By transforming the simulation results to a mesh with a defined number of mesh cells, distributions of 10,000 geometrical configurations were computed. By using Biot-Savarts law at virtual sensor positions, the x-components of the magnetic flux density were calculated. The simulated current density and $B_x(r)$ results for arrays of 10 x 10 and 50 x 50 virtual sensors in distances of 3, 5, 10 and 25 mm to the conductor serve as testing data set for various reconstruction methods. A model-based solution of an optimization problem is demonstrated, able to solve Biot-Savarts law inversely and to localize larger non-conductive fractions by reconstructing bendings of electric current streamline paths. Sensor arrays with a reasonably high number of sensors and distances smaller than 10 mm are needed to measure $B_x(r)$ with sufficient resolution for the streamline path optimization. An experimental validation of the numerically studied model is in preparation.

## REFERENCES

Angulo, A., van der Linde, P., Gardeniers, H., Modestino, M. and Fernández Rivas, D. (2020) 'Influence of Bubbles on the Energy Conversion Efficiency of Electrochemical Reactors.' Joule, 4(3) pp. 555–579.

Hauer, K.-H., Potthast, R., Wüster, T. and Stolten, D. (2005) 'Magnetotomography—a new method for analysing fuel cell performance and quality.' Journal of Power Sources, 143(1–2) pp. 67–74.

Kumar, N., Krause, L., Wondrak, T., Eckert, S., Eckert, K. and Gumhold, S. (2023) 'Learning to reconstruct the bubble distribution with conductivity maps using Invertible Neural Networks and Error Diffusion.' submitted to the 11th World Congress on Industrial Process Tomography, Mexico City.

Kwan, J. T. H., Nouri-Khorasani, A., Bonakdarpour, A., McClement, D. G., Afonso, G. and Wilkinson, D. P. (2022) 'Frequency Analysis of Water Electrolysis Current Fluctuations in a PEM Flow Cell: Insights into Bubble Nucleation and Detachment.' Journal of The Electrochemical Society, 169(5) p. 054531.

Panchenko, O., Borgardt, E., Zwaygardt, W., Hackemüller, F. J., Bram, M., Kardjilov, N., Arlt, T., Manke, I., Müller, M., Stolten, D. and Lehnert, W. (2018) 'In-situ two-phase flow investigation of different porous transport layer for a polymer electrolyte membrane (PEM) electrolyzer with neutron spectroscopy.' Journal of Power Sources, 390, June, pp. 108–115.

Plevachuk, Y., Sklyarchuk, V., Eckert, S., Gerbeth, G. and Novakovic, R. (2014) 'Thermophysical Properties of the Liquid Ga–In–Sn Eutectic Alloy.' Journal of Chemical & Engineering Data, 59(3) pp. 757–763.

Satjaritanun, P., O'Brien, M., Kulkarni, D., Shimpalee, S., Capuano, C., Ayers, K. E., Danilovic, N., Parkinson, D. Y. and Zenyuk, I. V. (2020) 'Observation of Preferential Pathways for Oxygen Removal through Porous Transport Layers of Polymer Electrolyte Water Electrolyzers.' iScience, 23(12) p. 101783.

Schmidtpeter, J., Wondrak, T., Makarov, D., Zabila, Y. and Sieger, M. (2023) 'Magnetic Particle Tracking enabled by Planar Hall Effect magnetic field sensors.' submitted to the 11th World Congress on Industrial Process Tomography, Mexico City.

Shepard, D. (1968) 'A two-dimensional interpolation function for irregularly-spaced data.' In Proceedings of the 1968 23rd ACM national conference on -. Not Known: ACM Press, pp. 517–524.

Yuan, S., Zhao, C., Cai, X., An, L., Shen, S., Yan, X. and Zhang, J. (2023) 'Bubble evolution and transport in PEM water electrolysis: Mechanism, impact, and management.' Progress in Energy and Combustion Science, 96, May, p. 101075.

Zhang, H. Q., Jin, Y. and Qiu, Y. (2015) 'The optical and electrical characteristics of PMMA film prepared by spin coating method.' IOP Conference Series: Materials Science and Engineering, 87, July, p. 012032.